\documentclass[conference,a4paper,top=0.7in]{IEEEtran}
\IEEEoverridecommandlockouts
\usepackage{cite}
\usepackage{amsmath,amssymb,amsfonts}
\usepackage{algorithm,algpseudocode}
\usepackage{graphicx}
\usepackage{textcomp}
\usepackage{xcolor}
\usepackage{enumitem}
\usepackage{resizegather}

\usepackage{stfloats}
\usepackage{accents}
\newlength{\dhatheight}

\def\BibTeX{{\rm B\kern-.05em{\sc i\kern-.025em b}\kern-.08em
    T\kern-.1667em\lower.7ex\hbox{E}\kern-.125emX}}
\usepackage{tikz}
\usepackage{mathtools}
\DeclareMathOperator{\Transpose}{T}
\newcommand{\Tr}{{\Transpose}}

\newcommand{\beq}{\begin{equation}}
\newcommand{\eeq}{\end{equation}}

\newcommand{\btau}{\boldsymbol{\tau}} 

\def\adots{\mathinner{\mskip0mu\raise0pt\vbox{\kern7pt\hbox{.}}\mskip3mu
          \raise4pt\hbox{.}\mskip3mu\raise8pt\hbox{.}\mskip0mu}}

\newcommand{\diag}{\mbox{diag}}

\newcommand{\alphah}{\widehat{\alpha}}

\newcommand{\bmh}{\bfh}

\usepackage{bm}

\newcommand{\bmx}{{\bm x}}
\newcommand{\bmy}{{\bm y}}

\newcommand{\bmF}{{\bm F}}

\renewcommand{\bmh}{{\bm h}}

\newcommand{\bmA}{{\bm A}}

\newcommand{\bmI}{{\boldsymbol {I}}}

\newcommand{\bmn}{{\boldsymbol {n}}}

\newcommand{\bmJ}{{\boldsymbol {J}}}
\newcommand{\bbeta}{{\boldsymbol {\beta}}}
\newcommand{\bPsi}{{\boldsymbol {\Psi}}}
\newcommand{\bmz}{{\boldsymbol {z}}}

\newcommand{\bGamma}{\boldsymbol{\Gamma}}
\newcommand{\bSigma}{\boldsymbol{\Sigma}}
\newcommand{\bSigmah}{\widehat{\bSigma}}
\newcommand{\bOmega}{\boldsymbol{\Omega}}
\newcommand{\bOmegah}{\widehat{\bOmega}}

\newcommand\norm[1]{\left\lVert#1\right\rVert}

\newcommand{\bTheta}{\boldsymbol{\Theta}}
\newcommand{\btheta}{\boldsymbol{\theta}}
\newcommand{\balpha}{\boldsymbol{\alpha}}
\newcommand{\balphah}{\widehat{\balpha}}
\newcommand{\bgamma}{\boldsymbol{\gamma}}

\begin{document}

\title{Sensing aided Channel Estimation in Wideband Millimeter-Wave MIMO
Systems\vspace{-4mm}
}
\vspace{-4mm}\author{
\IEEEauthorblockN{Rakesh Mundlamuri\IEEEauthorrefmark{1}, Rajeev Gangula\IEEEauthorrefmark{2}, Christo Kurisummoottil Thomas\IEEEauthorrefmark{3}, Florian Kaltenberger\IEEEauthorrefmark{1} and Walid Saad\IEEEauthorrefmark{3}
}
\IEEEauthorblockN{\IEEEauthorrefmark{1}Communication Systems Department,
EURECOM, Biot, France
}

\IEEEauthorblockN{\IEEEauthorrefmark{2}Institute for the Wireless Internet of Things, Northeastern University, Boston, USA 
}

\IEEEauthorblockN{
\IEEEauthorrefmark{3} Wireless@VT, Bradley Department of Electrical and Computer Engineering,
Virginia Tech, Arlington, VA, USA \\ 
}}
\maketitle

\begin{abstract}

In this work, the uplink channel estimation problem is considered for a millimeter wave (mmWave) multi-input multi-output (MIMO) system. It is well known that pilot overhead and computation complexity in estimating the channel increases with the number of antennas and the bandwidth. To overcome this, the proposed approach allows the channel estimation at the base station to be aided by the sensing information. The sensing information contains an estimate of scatterers locations in an environment. A simultaneous weighting orthogonal matching pursuit (SWOMP) - sparse Bayesian learning (SBL) algorithm is proposed that efficiently incorporates this sensing information in the communication channel estimation procedure. The proposed framework can cope with scenarios where a) scatterers present in the sensing information are not associated with the communication channel and b) imperfections in the scatterers' location. Simulation results show that the proposed sensing aided channel estimation algorithm can obtain good wideband performance only at the cost of fractional pilot overhead. Finally, the Cramer-Rao Bound (CRB) for the angle estimation and multipath channel gains in the SBL is derived, providing valuable insights into the local identifiability of the proposed algorithms.

\end{abstract}

\section{Introduction}

Millimeter wave (mmWave) and terahertz (THz) frequencies are considered to be a key component of 5G and 6G cellular systems\cite{saad2019vision}. However, as the operating frequencies increase, path and absorption losses also increase. Despite these disadvantages, this approach will allow packing more antennas in a small area and, then, the network can leverage beamforming techniques to compensate for the losses operating in such frequencies. However, the gains stemming from these multiple antenna techniques hinge on the ability to accurately estimate the channel state information (CSI).  

Estimating channel coefficients over a wideband and across multiple antennas incurs significant resource overhead in terms of resources occupied for sending pilot symbols. However, it has been observed that the mmWave channel exhibits a sparse behavior with only a few resolvable multi-paths in angle and delay domain \cite{RanRapErk_2014} and \cite{AkdRizEtal_2014}. By leveraging such sparsity, several works have come with compressed sensing (CS) based approaches for channel estimation and precoder design in mmWave multi-input multi-output (MIMO) systems\cite{SchniAkb_2014,AlkElLeuHeath_2014,LeeGilYong_2016,venugopal2017channel,8398433,8227682}. However, while used in wideband massive MIMO systems, these approaches lead to higher complexity due to the requirement of inverting huge matrices (for every subcarrier) across such antenna arrays. 
\begin{figure}[t]
\centerline{\includegraphics[width=3.2in,height=1.9in]{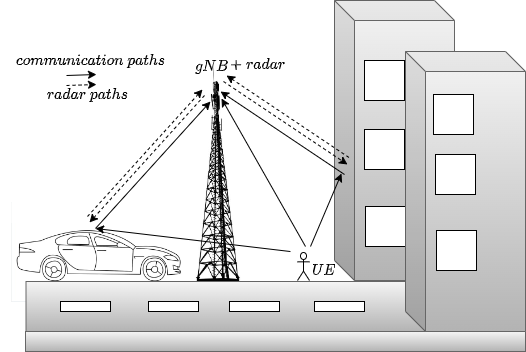}}
\vspace{-3mm}\caption{Uplink multi-path scenario along with the co-located radar.}
\label{fig:graphic_sec}\vspace{-7mm}
\end{figure}

Since the sparse wireless channel is described by a few geometric multi-path propagation parameters, one might ask: Can the information on the physical propagation environment, for example, scatter or reflector locations, be useful in channel estimation? Indeed, one of the earlier works in \cite{VasDeeKatabi_2016} has utilized this key observation. The authors extract physical multi-path parameters from the CSI measurements in  
one frequency band and then use them to construct the CSI in another frequency band. However, no extra pilots are used in aiding the channel estimation, and they assume that the extracted multi-path parameters are perfect.

On the other hand, advances in radar and joint communication sensing made it possible to have real-time dynamic radio environment maps at the communicating devices. Prior works in \cite{AliGonHeath_2018,ali2019millimeter,liu2020radar,WangNarHeath_2018,LaNuriaHeath_2019,chaccour2021joint} use such radar environment side information for the beam prediction and beam alignment to reduce the initial synchronization time in vehicular systems. A recent work in \cite{JiangAlkhateeb_2022} tried to address the problem of channel estimation in massive MIMO systems by leveraging radar sensing information. The authors retrieve the scatterers location and velocity in the surrounding environment from the measurements collected by a co-located radar at the base station (gNB). Then multi-path parameters such as delays and angles are extracted from the radar sensing information. These inferred multi-path parameters from the radar are then used to initialize the dictionary in an orthogonal matching pursuit (OMP) based channel estimation algorithm. However, the extracted multi-path parameters from the radar are assumed to be error-free. The limitations of existing literature on radar-based sensing for channel estimation in massive MIMO systems include the following. Firstly, the sensing based channel estimation algorithms may have limited angle 
 and delay resolution,
resulting in imprecise channel estimation and affecting the system
performance. Secondly, utilizing separate spectrum for sensing and communication leads to inefficient resource utilization. But this can be overcome by using full-duplex techniques \cite{HassaniSJ2020}, which requires efficient signal processing to mitigate the self-interference components. Thirdly, implementing sensing baded channel estimation in practice
can be difficult due to the increasing radio frequency (RF) components and computational complexity, especially in massive MIMO systems with
many antennas. These limitations highlight the need for further
research to improve sensing based channel estimation’s accuracy,
efficiency, and robustness in massive MIMO systems. Specifically, this paper proposes novel signal processing methods to bridge the gap on the limited angular and delay resolution issue mentioned above.

In this work, we consider the problem of channel estimation in a wideband mmWave MIMO system in which sensing information is obtained from a co-located radar at the gNB as shown in Fig.~\ref{fig:graphic_sec} is used to reduce the pilot overhead. The contributions of our paper are summarized as follows,
\begin{enumerate}
    \item Unlike \cite{JiangAlkhateeb_2022}, we assume that the sensing information from the radar can be erroneous. We also consider cases in which scatterers detected from radar might not be associated with the communication channel.
    \item To address these issues, a novel Simultaneous Weighting Orthogonal Matching Pursuit (SWOMP) - Sparse Bayesian Learning (SBL) based channel estimation is proposed that incorporates the imperfect sensing information from the radar.
    \item We also provide local identifiability analysis for the parameter estimation using SBL by deriving Cramer-Rao bound (CRB) for the joint angle of arrival (AoA) and path gain estimation using SBL.
\end{enumerate}

\section{System Model}

\begin{figure}[t]\vspace{-1mm}
\centerline{\includegraphics[width=3.5in]{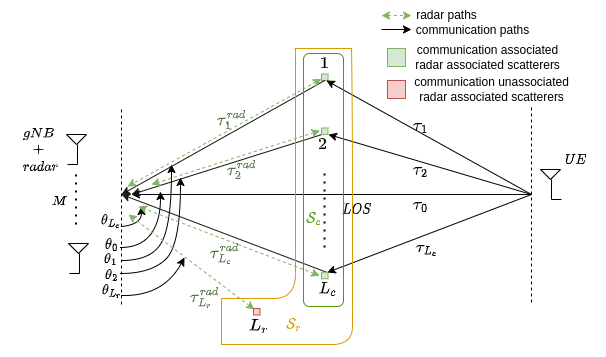}}
\vspace{-3mm}\caption{\hspace{-1mm}Scatterer environment along with the sensing information.}
\label{fig:channel_model_rad}
\end{figure}

We consider a scenario where a user (UE) communicates with base station (gNB) in an environment with the scatterers located between them. The scatterers are represented by $\mathcal{S}_r$, and $|\mathcal{S}_r|=L_r$. Only a subset of these scatterers, $\mathcal{S}_c \subseteq\mathcal{S}_r$, $|\mathcal{S}_c|=L_c$, are assumed to affect the UE-gNB communication channel.
The set $\mathcal{S}_c$ is unknown, however, we assume that location estimates of scatterers in $\mathcal{S}_r$ are provided by a sensing system.
This represents a scenario where the scatterers are present in the blind zone to UE but can be detected by a sensing system co-located at the gNB as shown in the Fig.~\ref{fig:channel_model_rad}. 

\subsection{Sensing Information}
We assume that the sensing is accomplished at the gNB either using a co-located radar operating in a seperate spectrum \cite{rao2017introduction} or through a joint communication sensing framework \cite{zhang2021overview}.
The sensing information available at the gNB is given by
$$
\{(\tau_{\ell}^{rad}+e_{\tau},\theta_l+e_{\theta}) ~ | ~ l=1,2,\ldots, L_r\},
$$
where $\tau_{\ell}^{rad}$ and $\theta_\ell$ represent the round trip delay and angle of the $l$-th scatterer from the gNB, respectively. The error in the estimated parameters is assumed to be Gaussian distributed as
$e_{\theta} \sim\mathcal{N}(0, \sigma_{\theta}^2)$ and 
$e_{\tau} \sim\mathcal{N}(0, \sigma_{\tau}^2)$.
The error in the radar spatial information can appear due to noise and the inability of the radar to resolve delay and/or angles sufficiently.

\subsection{Communication Model}

We consider a mmWave orthogonal frequency division multiplexing (OFDM) system with a single antenna UE and $M$ antenna gNB.
The gNB is equipped with a uniform linear array (ULA) with half-wavelength spacing between consecutive antennas.

The UE sends $P\ll K$ (narrowband) pilots, where $K$ is the total number of subcarriers used for communication. The received complex baseband signal at the $k$-th subcarrier after down-conversion, zero prefix removal, OFDM demodulation, and correlation with the pilots is given by
\begin{align}
\boldsymbol{y}[k] = \boldsymbol{h}[k] + \boldsymbol{n}[k],
\label{eq_rx_sig}
\end{align}
where $k=0$ to $K-1$, $\boldsymbol{h}[k]\in\mathbb{C}^{M\times 1}$ represent the baseband channel, 
$\boldsymbol{n}[k]\sim\mathcal{CN}(0, \sigma^2\boldsymbol{I}_{M})$ is a circularly symmetric complex Gaussian distributed additive noise vector. We define the received signal-to-noise-ratio (SNR) at subcarrier $k$ as ${\norm{\bmh[k]}^2}/{\sigma^2}$. Next, we describe the mmWave channel model generation that is a parametric function of the multipath components.
\subsection{Channel Model}\label{sec:ch_mod}
A frequency-selective geometric channel model with $N_c$ delay taps and $L_c+1$ paths\cite{venugopal2017channel} is considered. The channel consists of a line-of-sight (LoS) component, and $L_c$ (yet unknown) reflections resulting from the scatterers as described earlier. 
The $d$-th delay tap is modeled as 
\begin{align} 
  \boldsymbol{h}_d = \sqrt{\frac{M}{L_c+1}}\sum\limits_{\ell=0}^{L_c} \alpha_\ell p(dT_s-\tau_\ell) \boldsymbol{a}(\theta_\ell), \label{eq:Hd}
\end{align}
where 
$p(.)$ is the pulse-shaping filter, $T_s$ is the sampling interval, 
$\alpha_\ell$, $\tau_\ell$, $\theta_\ell$ represent the path gain, delay and the angle-of-arrival (AoA) of the $l$-th path, respectively. The receiver array steering vector for the $l$-th path is denoted by $\boldsymbol{a}(\theta_\ell) \in \mathbb{C}^{M\times 1}$.   
The index $\ell{=}0$ is always associated with the LoS path. We can compactly represent the channel as
 $\boldsymbol{h}_d{=} \boldsymbol{A}\boldsymbol{\Delta}_d,$
where $ \boldsymbol{A}{=} [\boldsymbol{a}(\theta_0)\  \boldsymbol{a}(\theta_1)\  \dots\ \boldsymbol{a}(\theta_{L_c})]\in \mathbb{C}^{M\times \left(L_c+1\right)}$ contains the receiver side steering vectors and
\begin{align}
\boldsymbol{{\Delta}}_d =  \left[{\begin{array}{c} \alpha_0 p (dT_s-\tau_0), \cdots,  \alpha_{L_c} p (dT_s-\tau_{L_c})
\end{array}}\right]^{\Tr}. 
\label{delta_d_m}
\end{align}
 We obtain the frequency domain channel representation by taking a $K$-point DFT of the delay-domain channel, and the channel at subcarrier $k$ can be written as
\begin{align} 
 \boldsymbol{h}[k] = \sum\limits_{d=0}^{N_c-1}  \boldsymbol{h}_d \exp\left({-\frac{j2\pi kd}{K}}\right) =  \boldsymbol{A} \boldsymbol{{\Delta}}[k], \label{eq:Hk}
\end{align} 
and $\boldsymbol{{\Delta}}[k]$ is given by
$
\boldsymbol{{\Delta}}[k] = \sum\limits_{d=0}^{N_c-1}\boldsymbol{{\Delta}}_d \exp\left({-\frac{j2\pi kd}{K}}\right).$
Further substituting for $\boldsymbol{{\Delta}}_d$ from \eqref{delta_d_m}, we obtain
\begin{align}
\boldsymbol{\Delta}[k]=\left[{\beta_{k,0}\alpha_0},{\beta_{k,1}\alpha_1}, \dots,  {\beta_{k,L_c}\alpha_{L_c}}\right]^{\Tr},
\end{align}
where $\beta_{k,\ell}{=}\sum
_{d=0}^{N_c-1}p(dT_s-\tau_\ell)\exp\left({-\frac{j2\pi kd}{K}}\right)$. Substituting $\boldsymbol{{\Delta}}[k]$ in \eqref{eq:Hk}, a compact form of the frequency domain channel $\boldsymbol{h}[k]$ can be obtained as
\begin{align}
    \boldsymbol{h}[k] =\boldsymbol{A}\boldsymbol{\beta}_k\boldsymbol{\alpha},
\end{align}
where $\boldsymbol{\beta}_k{=}\textrm{diag}\big(\beta_{k,0},\beta_{k,1},\dots,\beta_{k,{L_c-1}}\big)$, and $\boldsymbol{\alpha}{=}[\alpha_0,\alpha_1,\dots,\alpha_{L_c}]^{\Tr}$. Further substituting $\boldsymbol{h}[k]$ in \eqref{eq_rx_sig}, the received frequency domain signal $\boldsymbol{y}[k]$ can be written as
\begin{align}
    \boldsymbol{y}[k]=\boldsymbol{\Psi}_k\boldsymbol{\alpha} + \boldsymbol{n}[k],
    \label{eq_underscore_dict}
\end{align}
where $\boldsymbol{\Psi}_k = \boldsymbol{A}\boldsymbol{\beta}_k\in \mathbb{C}^{M\times \left(L_c+1\right)}$.

\section{Sensing Aided Channel Estimation}
In this section, we provide a channel estimation framework that incorporates the sensing information available at the gNB. From \eqref{eq_underscore_dict}, the received $P$ pilots in vectorized form is given by
\begin{align}
    & \boldsymbol{y} = \left[\boldsymbol{y}^{\Tr}[0] \,\boldsymbol{y}^{\Tr}[1]\, \ldots \, \boldsymbol {y}^{\Tr}[P-1] \right]^{\Tr},\\
    & \boldsymbol{y} = \underbrace{\left[ {\boldsymbol{\Psi}_0^{\Tr}} {\boldsymbol{\Psi}_1^{\Tr}} \cdots {\boldsymbol{\Psi}_{P-1}^{\Tr}}\right]^{\Tr}}_{{\bOmega}}\boldsymbol{\alpha} + \bmn, 
    \label{eq:omega}
\end{align}
where the matrix $\bOmega\in \mathbb{C}^{MP\times (L_c+1)}$ carries the delay-angle information of the multipath components and $\bmn$ is the vectorized noise $\bmn=[\bmn^{\Tr}[0] \bmn^{\Tr}[1] \cdots \bmn^{\Tr}[P-1]]^{\Tr}$.

\begin{figure}[t]\vspace{-2mm}
\centerline{\includegraphics[width=3.5in,height=1.21in]{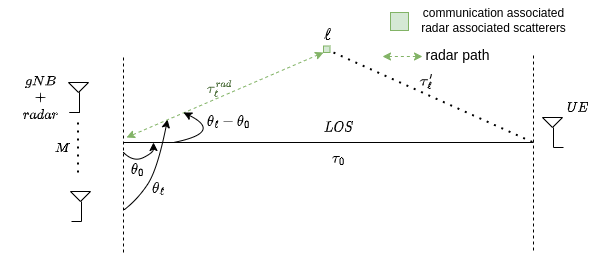}}\vspace{-2mm}
\caption{Communication delay estimation from radar delay.}
\label{fig:channel_model_rad_delay_est}\vspace{-4mm}
\end{figure}

Moreover, the sensing information can be used as an initial estimate of the multipath delays and angles. Let $\tilde{\bm{\theta}}=[{\theta}_0,\tilde{\theta}_1, \tilde{\theta}_2,\ldots, \tilde{\theta}_{L_r}]$, where ${\theta}_0$ is the angle associated with the LoS path and $\tilde{\theta}_l$, $l \in [1,L_r]$ is the AoA of the $l$-th path obtained from the sensing information. The round-trip propagation delay between the gNB and the $l$-th, $l \in [1,L_r]$, scatterer is denoted by
$\tau^{rad}_{\ell}$.
Let us define $\tilde{\bm{\tau}} = 
[{\tau_0}, \tilde{\tau}_1,\tilde{\tau}_2,\ldots,\tilde{\tau}_{L_r}]$, where $\tau_0$ is the delay between the UE and the gNB, and 
the delay of the     
 $\ell$-th communication path can be estimated using the radar delay $\tau^{rad}_{\ell}$ as
\begin{equation}
    \tilde{\tau}_{\ell} = {\tau^{rad}_{\ell}}/{2} + \tau^{\prime}_{\ell},
    \label{eq:tau_ell}
\end{equation}
where $\tau^{\prime}_{\ell}$ is obtained using triangle laws of cosines as shown in Fig.~\ref{fig:channel_model_rad_delay_est},
\begin{align}
  \tau^{\prime}_{\ell} = \sqrt{\tau_{0}^2 + \left({\tau^{rad}_{\ell}}/{2}\right)^2 - \tau_{0}(\tau^{rad}_{\ell})cos(\tilde{\theta}_{\ell}-\theta_{0})}.
  \label{eq:tau_prime}
\end{align}

Similar to the matrix $\bOmega$ in \eqref{eq:omega}, using the sensing information $(\tilde{\bm{\tau}}, \tilde{\bm{\theta}})$,
we can construct a matrix $\bar{\bOmega}\in\mathbb{C}^{MP\times (L_r+1)}$ that captures the delay-angle information of the $L_r+1$ paths. 
As we described earlier, only a subset of $L_c$ among the $L_r$ scatterers are included in the communication channel, and $L_c$ is unknown. This can be mathematically represented as,  
\begin{equation}\vspace{-1mm}
    \bOmega = \bar{\bOmega}\boldsymbol{B} + \boldsymbol{E},
    \label{eq_omega_err}\vspace{-1mm}
\end{equation}
where $\boldsymbol{B}\in\mathbb{R}^{(L_r+1)\times (L_c+1)}$ is obtained by selecting $L_c+1$ columns of the identity matrix $\boldsymbol{I}_{L_r+1}$. The indices of the columns that are included in $\boldsymbol{B}$, correspond to the paths that are present both in the communication channel and sensing information.
The unknown error term is denoted by $\boldsymbol{E}$.
\vspace{-1mm}\subsection{Problem Formulation}
Utilizing the received pilot signal \eqref{eq:omega} and the sensing information in the form of \eqref{eq_omega_err}, 
%
the maximum a posteriori (MAP) based channel estimation problem is formulated as:
\begin{align}\vspace{-1mm}
\left[\bOmega^*, \balpha^*\right] = \arg \max\limits_{\bOmega,\balpha} p(\bOmega,\balpha\mid \boldsymbol{y}),
\label{eq_MAP}\vspace{-1mm}
\end{align}
where $p(.)$ represents the probability distribution and $\balpha$ is the channel gain vector.

The optimization problem at hand is difficult to solve in general as a) it is hard to obtain the distribution $p(\bOmega,\balpha\mid \boldsymbol{y})$ b) the combinatorial nature of the path association matrix $\boldsymbol{B}$ and the unknown error.
A conventional approach to relax this problem and
solve it using 
compressed sensing schemes, such as SBL, by considering a joint dictionary matrix consisting of finely spaced angles and delays. However, such a solution results in cubic complexity with respect to the dictionary dimensions, which has to be finely spaced to alleviate the off-grid errors. Hence, we propose a two-stage SWOMP-SBL algorithm to overcome such high complexity. 



\vspace{-2mm}\section{SWOMP-SBL Algorithm}
\begin{figure}[t]\vspace{-2mm}
\centerline{\includegraphics[width=4.5in]{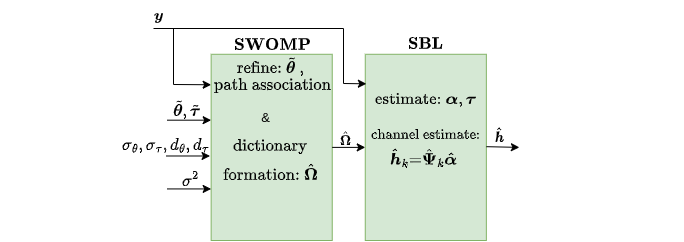}}
\vspace{-2mm}\caption{\hspace{-1mm} SWOMP-SBL algorithm.}\vspace{-5mm}
\label{fig:block_diagram}
\end{figure}

The proposed algorithm works in two stages. In the first stage, based on the sensing information, a SWOMP based algorithm is used to find the paths that are associated with the communication and their respective AoAs. Based on these selected paths, a SBL inference algorithm is used to obtain finer estimate of the delays and corresponding channel gains $\balphah$. A schematic describing this two-stage algorithm is shown in Fig \ref{fig:block_diagram}.

\vspace{-3mm}\subsection{SWOMP Stage}
The algorithm is initialized with assuming that
all the $L_r+1$ paths from the sensing information are present in the communication channel. The AoA's $\tilde{\btheta}$ are used to form the angle dictionary $\boldsymbol{A}^{\prime}$ as described in steps 2 and 3 of the Algorithm \ref{SWOMP}. 
The SWOMP algorithm \cite{rodriguez2018frequency} outputs the maximum correlated paths $\hat{\btheta}$ corresponding to the angle dictionary $\boldsymbol{A}^{\prime}$ with the received signal $\boldsymbol{y}$. The noise variance $\sigma^2$ is utilized as a stopping condition in SWOMP, where all the refined angles associated with the channel and their corresponding path indices $\boldsymbol{\chi}$ are estimated. However, a dictionary matrix $\bOmegah$ is needed to refine the delays further and estimate the channel gains. The path association matrix $\boldsymbol{B}$ can be obtained from the estimated $\boldsymbol{\chi}$, but it's avoided since the path indices are enough to create the dictionary matrix $\bOmegah$. $\bOmegah$ is constructed using the refined AoA $\hat{\btheta}$ obtained using SWOMP and a finely space dictionary matrix of the associated delays. The association of the path is given by the path indices $\boldsymbol{\chi}$ and maps the refined angles to their corresponding delays. The details of our algorithm are discussed in Algorithm~\ref{SWOMP}.
The refinement of the delays $\tilde{\btau}$ and their corresponding channel gains $\balpha$ are estimated using SBL with the obtained $\bOmegah$ in the next stage. The computational complexity of SWOMP in each iteration is $MP (d_\theta L^{\prime})^2 +MPd_{\theta}L_r + MPd_{\theta}L^{\prime}$.
\subsection{SBL Stage}
Recalling the measurement equation with the obtained $\bOmegah$, we write $\boldsymbol{y} = \bOmegah{{ {\boldsymbol{\alpha}}}} + {{{\boldsymbol {n}}}}$. We formulate the estimation method of $(\balpha,\tilde{\btau})$ using SBL as follows.

SBL is a type-II maximum likelihood (ML) estimation procedure to obtain the channel estimate~\cite{wipf2004sparse,zhang2011sparse}. In this method, ${\boldsymbol{\alpha}}$ is considered as a hidden variable, and posterior statistics are obtained given the observations. SBL assumes a complex Gaussian prior distribution for the entries of ${\boldsymbol{\alpha}}$, which gets written as $
p(\alpha_i){=}\frac{\gamma_i}{\pi}^{-\gamma_i(\alpha_i)^2}.$
The hyperparameters $\alpha_i$ also estimated using the inference procedure. $\gamma_i$ is assumed to follow a Gamma distribution, $\mathcal{G}(\gamma_i;a,b){=}\Gamma((\gamma_i))^{-1}(\gamma_i)^{a-1}e^{-b\gamma_i}b^{\gamma_i}$. Defining, $\bGamma{=}\diag(\bgamma)$, where $\bgamma$ is the vector of $\gamma_i$. Noise is assumed to be complex Gaussian, $\mathcal{CN}(\bf{0},\frac{1}{\zeta}\bmI)$. $\zeta$ is assumed to have Gamma as a prior distribution such that $ p_{\zeta}(\zeta){=}\mathcal{G}(\zeta;c,d)$, where $c,d$ are known. Note that in the case of an uninformative prior, the values of $a$ and $b$ corresponds to 1 and 0 respectively. Now, the posterior distribution of $\balpha$ and the hyper-parameters $\boldsymbol{\gamma}, \zeta$ needs to be obtained. Since the prior and the noise are both Gaussian, obtaining the posterior statistics of $\balpha$ is straightforward. But, the computation of $\boldsymbol{\gamma}$ requires the computation of the marginal probability distribution $p(\boldsymbol{y};\boldsymbol{\gamma},\zeta)$ and maximizing it (alternatively) w.r.t. $\boldsymbol{\gamma},\zeta$. This procedure is known as evidence maximization or type-II ML estimation.

To solve this, expectation-maximization (EM) algorithm is used, which proceeds by lower bounding the logarithm of the evidence $p(\boldsymbol{y};\boldsymbol{\gamma},\zeta)$, and maximizing it iteratively. Treating $\balpha$ as a hidden variable, In the expectation (E) step, expectation of the log likelihood of $(\boldsymbol{y},\balpha)$ w.r.t. $p(\balpha|\boldsymbol{y},\boldsymbol{\gamma},\zeta)$ is computed. In the maximization (M) step, the hyper-parameters $\boldsymbol{\gamma},\zeta$ are computed by maximizing the function obtained in the E step. More details of SBL and type-II ML estimation can be found in \cite{wipf2004sparse}. Detailed steps for the channel estimation are provided in Algorithm~\ref{SBL}. The SBL algorithm outputs the estimate of the channel gains $\balphah$. Using step 17 in Algorithm~\ref{SWOMP}, the channel estimate $\hat{\boldsymbol{h}}$ at the $k$-th subcarrier can be obtained by $\hat{\boldsymbol{h}}_k{=}\hat{\boldsymbol{\Psi}}_k\balphah$ for all the $K$ subcarriers.
The convergence properties of the SBL algorithm are well understood in the literature \cite{wipf2004sparse}. In short, using similar arguments in \cite{wipf2004sparse}, we can show that the proposed SBL converges to the sparsest solution when the noise variance is zero and to a sparse local minimum, irrespective of the noise variance. The computational complexity of each iteration of SBL is $(MP)^3+ 2(MP)^2(d_{\tau}L^{\prime})+2(d_{\tau}L^{\prime})^2MP+(d_{\tau}L^{\prime})^2+ 4MP(d_{\tau}L^{\prime})+(d_{\tau}L^{\prime})$.
\setlength{\textfloatsep}{0pt}
\vspace{0mm}\begin{algorithm}[t]
\caption{SWOMP Stage}\label{SWOMP}
\begin{algorithmic}[1]\vspace{-0mm}
\Require $\boldsymbol{y}, \tilde{\btheta}, \tilde{\btau}, d_{\theta}, d_{\tau}, \sigma_{\theta}, \sigma_{\tau}, \sigma^2$\\
Initialize: $\boldsymbol{\chi}=\{\}$
\vspace{1mm}
\State $\btheta_l^{\prime}=\tilde{\theta}_l-2\sigma_{\theta}:\frac{4\sigma_{\theta}}{d_{\theta}}:\tilde{\theta}_l+2\sigma_{\theta}\in\mathbb{R}^{1 \times d_{\theta}}$
\vspace{1mm}
\State $\boldsymbol{A}^{\prime}$=$[a(\btheta_0^{\prime})\ a(\btheta_1^{\prime})\ \cdots\ a(\btheta_{L_r}^{\prime})]\in\mathbb{C}^{M\times d_{\theta}(L_r+1)}$ 
\vspace{1mm}
\State $\hat{\btheta}=\textbf{SWOMP}(\boldsymbol{y},
\boldsymbol{A}^{\prime},\sigma^2)$ \\
\vspace{1mm}
$\hat{\btheta}\in\mathbb{R}^{1\times L^{\prime}}$ is $\{\hat{\theta}_{\ell}\ |\ \ell=0,1,\cdots,L^{\prime}-1\}$\\
\vspace{1mm}
$\textbf{Path association:}$
\For{$\ell=0:L^{\prime}-1$}
\State $p = \arg \min\limits\{|\theta_\ell\boldsymbol{1}-\btheta|\}$ \Comment{$\boldsymbol{1}\in 1^{1\times (L_r+1)}$}
\State $\boldsymbol{\chi}=\boldsymbol{\chi} \cup p$ \Comment{$p{\ =}\ $path index}
\EndFor
\State $\tilde{\btau}(\boldsymbol{\chi})\in\mathbb{R}^{1\times L^{\prime}}$ is $\{\tilde{\tau}_{\ell}(\boldsymbol{\chi})\ |\ \ell=0,1,\cdots,L^{\prime}-1\}$ \Comment{\!delays of the corresponding path index obtained in step 9}
\vspace{1mm}
\State $\hat{\btau}_{\ell}=\tilde{\tau}_{\ell}(\boldsymbol{\chi})-2\sigma_{\tau}:\frac{4\sigma_{\tau}}{d_{\tau}}:\tilde{\tau}_{\ell}(\boldsymbol{\chi})+2\sigma_{\tau}\in\mathbb{R}^{1 \times d_{\tau}}$
\vspace{1mm}
\State $\hat{\btau} = [\hat{\btau}_0\ \hat{\btau}_1\ \cdots\ \hat{\btau}_{L^\prime-1}]\in\mathbb{R}^{1 \times d_{\tau}L^{\prime}}$
\vspace{1mm}
\State The resulting $\boldsymbol{\beta}_{k}$ obtained using $\hat{\btau}$ is denoted as  $\hat{\boldsymbol{\beta}}_{k}\in\mathbb{C}^{d_{\tau}L^{\prime} \times d_{\tau}L^{\prime}}$
\vspace{1mm}
\State $\hat{\boldsymbol{A}}_{\ell}=[a(\hat{\theta}_{\ell})a(\hat{\theta}_{\ell})\cdots a(\hat{\theta}_{\ell})]\in\mathbb{C}^{M \times d_{\tau}}$\Comment{Repeat $d_{\tau}$ times}
\vspace{1mm}
\State $\hat{\boldsymbol{A}}=[\hat{\boldsymbol{A}}_0\ \hat{\boldsymbol{A}}_1\ \cdots\ \hat{\boldsymbol{A}}_{L^{\prime}-1}]\in\mathbb{C}^{M \times d_{\tau}L^{\prime}}$
\vspace{1mm}
\State $\hat{\boldsymbol{\Psi}}_{k} = \hat{\boldsymbol{A}}\hat{\boldsymbol{\beta}}_{k}\in\mathbb{C}^{M \times d_{\tau}L^{\prime}}$
\vspace{1mm}
\State $\bOmegah=[\hat{\boldsymbol{\Psi}}_0^{\Tr} \hat{\boldsymbol{\Psi}}_1^{\Tr} \cdots \hat{\boldsymbol{\Psi}}_{{P-1}}^{\Tr}]^{\Tr}\!\in\!\mathbb{C}^{MP \!\times d_{\tau}L^{\prime}}$
\end{algorithmic}\vspace{-1.5mm}
\vspace{-0mm}
\end{algorithm}\vspace{-1mm}
\setlength{\textfloatsep}{0pt}
\begin{algorithm}[t]
\caption{SBL Stage}\label{SBL}
\begin{algorithmic}[1]
\Require $\boldsymbol{y}$, $\bOmegah$,  $p_{\boldsymbol{y}}(\boldsymbol{y}\mid\bOmegah,\boldsymbol{\alpha} )$ \\
Initialize: $t=0$, $(\balphah)^t$ using LS estimate.
\Repeat
\State [Estimate $\balpha$]
\vspace{1mm}
\State $\bSigma_y^t = \frac{1}{\zeta^{t-1}}\bmI + \bOmegah\bGamma^{-1}\bOmegah^H$.
\vspace{1mm}
\State $\bSigmah^t = \bGamma^{-1} - \bGamma^{-1}\bOmegah^H\left(\bSigma_y^t\right)^{-1}\bOmegah\bGamma^{-1}$.
\vspace{1mm}
\State $(\balphah)^t = \zeta^{t-1}\bSigmah^{t}\bOmegah^H\boldsymbol{y}$.
\vspace{1mm}
\State [Hyper-parameters Update]
\vspace{1mm}
\State $\gamma_i^t = \frac{2a-1}{\mathbb{E}\left((\alphah_i)^2\right)+2b}$.
\vspace{1mm}
\State $\zeta^t = \frac{2c-1}{\frac{\mathbb{E}(\lvert \boldsymbol{y}-\bmz \rvert^2)}{MP}+2c}, z_i = \bOmegah \balphah$.
\Until {Convergence}
\end{algorithmic}
\end{algorithm}

\vspace{-2mm}\subsection{Identifiability of the Proposed SBL: Minimum Narrowband Pilots Required?}
\vspace{-2mm}

This subsection provides conditions under which the sensing-aided channel estimation using SBL becomes locally identifiable. Furthermore, the analysis herein provides the minimum pilots required for the respective channel estimation algorithm to be identifiable. The distribution of $\boldsymbol{y}[k]$, after marginalizing w.r.t $\balpha$ can be written as
\vspace{-1mm}\begin{align}
p_{\boldsymbol{y}}(\boldsymbol{y}[k]) = \mathcal{CN}(\boldsymbol{0}, \bPsi_k \bGamma^{-1}(\bPsi_k)^H + \zeta^{-1}\bmI).\vspace{-2mm}
\end{align} 
The signal model in  \eqref{eq_underscore_dict} is non-identifiable if $\bPsi_k \bGamma^{-1}_1(\bPsi_k)^H{=}\bPsi_k \bGamma^{-1}_2(\bPsi_k)^H$ for some $\bGamma^{-1}_1 \neq \bGamma^{-1}_2$. The rank of $\bPsi_k^{\Tr}\otimes \bPsi$ is denoted as $R \leq \left(MP\right)$, where $\otimes$ represents the Khatri-Rao product. Following similar analysis\cite{PalICASSP2014}, we can show that the SBL algorithm is identifiable as long as $L$  (number of nonzero elements in $\balpha$) is $\mathcal{O}(R^2)$ (${=}\mathcal{O}(\left(MP\right)^2)$ for suitable $\bPsi_k^{\Tr}$). For a mmWave system, this would be just fewer pilots, compared to using number of pilots of the $\mathcal{O}(K)$ as in existing 5G-NR algorithms. 

Next, we look at the CRB of the estimation model here. The local identifiability (upto permutation ambiguity) of the SBL based parameter estimation is ensured if the Fisher information matrix (FIM) is non-singular \cite{BoizardICASSP2015}. First, the estimated parameters are defined in a vector as $\bTheta {=}[\btheta, \balpha, \bgamma, \zeta, \boldsymbol{\tau}]$. The FIM can be partitioned as
\vspace{-1mm}\begin{align}\vspace{-2mm}
\small \bmJ_{\bTheta\bTheta} = \begin{bmatrix}
\bmJ_{\btheta\btheta} &\bmJ_{\btheta\balpha} &\bmJ_{\btheta\bgamma} &\bmJ_{\btheta\zeta} &\bmJ_{\btheta\btau}\\
\bmJ_{\balpha\btheta} &\bmJ_{\balpha\balpha} &\bmJ_{\balpha\bgamma}& \bmJ_{\balpha\zeta} & \bmJ_{\balpha\btau}\\ 
\bmJ_{\bgamma\btheta} &\bmJ_{\bgamma\balpha} &\bmJ_{\bgamma\bgamma}& \bmJ_{\bgamma\zeta} & \bmJ_{\bgamma\btau} \\ 
\bmJ_{\zeta\btheta} &\bmJ_{\zeta\balpha} &\bmJ_{\zeta\bgamma}& \bmJ_{\zeta\zeta}& \bmJ_{\zeta\btau}, \\ 
\bmJ_{\btau\btheta} &\bmJ_{\btau\balpha} &\bmJ_{\btau\bgamma}& \bmJ_{\btau\zeta}& \bmJ_{\btau\btau}
\end{bmatrix},\vspace{-5mm}
\end{align}\vspace{-1mm}
where $\bmJ_{\bmx\bmy}{=} \mathbb{E}\left(\frac{\partial \ln p(\bmy,\bmx)}{\partial \bmx}  \frac{\partial \ln p(\bmy,\bmx)}{\partial \bmy}^{\Tr}\right)$. Each of the FIM blocks can be derived as (detailed derivations are skipped since those follows classical results in estimation theory)
\begin{align}\vspace{-4mm}
    \bmJ_{\btheta\btheta} &= \mathbb{E}(\zeta)(\bbeta_k)^H\mathbb{E}(\frac{\partial \bmA(\btheta)}{\partial \btheta}^H\frac{\partial \bmA(\btheta)}{\partial \btheta})\bbeta_k\mathbb{E}(\bGamma^{-1}) , \\
    \bmJ_{\btheta\zeta} &= \diag\left(\Re\{(\bbeta_k)^H\frac{\partial \bmA(\btheta)}{\partial \btheta}^H\bmA\bbeta_k\}\mathbb{E}(\bGamma^{-1})\right), \\
     \bmJ_{\bgamma\bgamma} &= -\mathbb{E}(\bGamma^{-1}) + (a-1)\mathbb{E}(\bGamma^{-1}), \\ \bmJ_{\zeta\zeta} &= -MP\mathbb{E}(\zeta^{-2}) + (c-1)\mathbb{E}(\zeta^{-1}),  \\
         \bmJ_{\balpha\balpha} &= -\mathbb{E}(\bGamma)- (\bbeta_k)^H(\bmA)^H\bmA\bbeta_k\mathbb{E}(\zeta),  \\ \bmJ_{\zeta\btau} &= \frac{\partial(\bbeta_k)^H}{\partial \btau}\mathbb{E}\left( \bmA(\btheta)^H\bmA(\btheta)\right)\bbeta_k\mathbb{E}(\bGamma^{-1}) , \\  \bmJ_{\btau\btau} &= \mathbb{E}(\zeta)\frac{\partial(\bbeta_k)^H}{\partial \btau}\mathbb{E}\left( \bmA(\btheta)^H\bmA(\btheta)\right)\frac{\partial\bbeta_k}{\partial \btau}\mathbb{E}(\bGamma^{-1}) , \vspace{-4mm}
    \end{align}\vspace{-4mm}\begin{align}\vspace{-3mm}
           \bmJ_{\btheta\btau} &= \mathbb{E}(\zeta)\frac{\partial(\bbeta_k)^H}{\partial \btau}\mathbb{E}\left( \bmA(\btheta)^H\frac{\partial\bmA(\btheta)}{\partial \btheta}\right)\bbeta_k\mathbb{E}(\bGamma^{-1}) , 
         \vspace{-5mm}
\end{align}  \vspace{-2mm}
and rest of the terms result to be zero. 
\begin{align}\vspace{-3mm}\small
\bmJ_{\bTheta\bTheta} = \begin{bmatrix}
\bmJ_{\btheta\btheta} & \boldsymbol{0} &\boldsymbol{0} &\bmJ_{\btheta\zeta} & \bmJ_{\btheta\btau} \\
\boldsymbol{0} &\bmJ_{\balpha\balpha} &\boldsymbol{0}& \boldsymbol{0} &\boldsymbol{0}\\ 
\boldsymbol{0} &\boldsymbol{0} &\bmJ_{\bgamma\bgamma}& \boldsymbol{0}&\boldsymbol{0} \\ 
\bmJ_{\zeta\btheta} &\boldsymbol{0} & \boldsymbol{0}& \bmJ_{\zeta\zeta} & \bmJ_{\zeta\btau} \\ 
\bmJ_{\btau\btheta} &\boldsymbol{0} & \boldsymbol{0}& \bmJ_{\btau\zeta} & \bmJ_{\btau\btau} 
\end{bmatrix}.\vspace{-2mm}
\end{align}
\begin{figure*}
\begin{equation}\tag{27}\begin{aligned}\vspace{-4mm}
&CRB(\btheta,\balpha) =  \left(\begin{bmatrix}\bmJ_{\btheta\btheta} & \boldsymbol{0}\nonumber \\ \boldsymbol{0} & \bmJ_{\balpha\balpha}\end{bmatrix} -  \begin{bmatrix}\tiny\boldsymbol{0} & \bmJ_{\btheta\zeta} & \bmJ_{\btheta\btau} \\ \boldsymbol{0} & \boldsymbol{0} & \boldsymbol{0}\end{bmatrix}\begin{bmatrix}\tiny\bmJ_{\bgamma\bgamma}^{-1} & \boldsymbol{0} \\ \boldsymbol{0} & \underbrace{\begin{bmatrix}\bmJ_{\zeta\zeta} & \bmJ_{\zeta\btau} \\
\bmJ_{\btau\zeta} & \bmJ_{\btau\btau}\end{bmatrix}^{-1}}_{\bmF_{\zeta,\btau}}\end{bmatrix}\begin{bmatrix}\tiny\boldsymbol{0} & \bmJ_{\btheta\zeta} & \bmJ_{\btheta\btau}\\ \boldsymbol{0} & \boldsymbol{0} & \boldsymbol{0}\end{bmatrix}^{\Tr}\right)^{-1},\vspace{-2mm}
\end{aligned}\label{eq_CRB_1}
\end{equation}\vspace{-6mm}
\end{figure*}The CRB for $\bTheta$ can be expressed as $CRB(\bTheta) {=}\bmJ_{\bTheta\bTheta}^{-1}$. The CRB for AoA estimates and $\balpha$ can be written using Schur-complement for inverting a block matrix as derived in \eqref{eq_CRB_1}, 
which can be simplified as
\vspace{-2mm}\begin{align}\vspace{-2mm}\scriptsize
&CRB(\btheta,\balpha) = \\\nonumber & \begin{bmatrix}\left(\bmJ_{\btheta\btheta}-\begin{bmatrix}
    \bmJ_{\btheta\zeta}&\bmJ_{\btheta\btau}
\end{bmatrix}\bmF_{\zeta\btau}^{-1}\begin{bmatrix}
    \bmJ_{\btheta\zeta}&\bmJ_{\btheta\btau}
\end{bmatrix}^{\Tr}\right)^{-1}  & \boldsymbol{0} \\ \boldsymbol{0} & \bmJ_{\balpha\balpha}^{-1}\end{bmatrix} .
\label{eq_crb_theta_alpha}\vspace{-6mm}
\end{align}
Following similar derivations, the CRB can be computed for $\balpha,\bgamma,\zeta$. From (26), we can conclude that for local identifiability of $\btheta,\balpha$, $\bmJ_{\btheta\btheta}-\begin{bmatrix}
    \bmJ_{\btheta\zeta}&\bmJ_{\btheta\btau}
\end{bmatrix}\bmF_{\zeta\zeta}^{-1}\begin{bmatrix}
    \bmJ_{\btheta\zeta}&\bmJ_{\btheta\btau}
\end{bmatrix}^{\Tr}$, and  $\bmJ_{\balpha\balpha}$ should be invertible, respectively.

\section{Simulation Results}

\begin{figure}[t]
\centerline{\includegraphics[width=3in,height=0.8in]{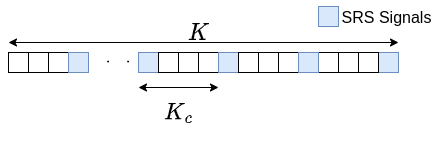}}\vspace{-4mm}
\caption{Uplink SRS comb structure in a OFDM symbol.}
\label{fig:comb}
\end{figure}

In this section, the performance of our novel SWOMP-SBL sensing aided channel estimation algorithm is evaluated through numerical simulations. The system parameters considered here are,  subcarrier spacing${=}120 \,\textrm{KHz}$,  center frequency${=}28 \,\textrm{GHz}$, sampling rate${=}30.72 \,\textrm{MHz}$, sampling period $T_s{=}32.552\, \textrm{ns}$, fft size $K{=}256$, cyclic prefix length $N_{cp}{=}34$ and the number of receive antennas $M{=}32$.
The pilots are generated similar to the sounding reference signals (SRS) in 5G standards \cite{3GPP5GNR}. The location of the pilots in the OFDM grid are arranged in a comb fashion as defined in the 3GPP standard\cite{3GPP5GNR} i.e., one pilot for every $K_c$ subcarriers as shown in Fig.~\ref{fig:comb}. The channel is generated using a ray-tracing tool using the locations of gNB and UE with the number of delay taps $N_c{=}N_{cp}$. The erroneous sensing information is generated with $\sigma_{\theta}{=}3^{\circ}$ and $\sigma_{\tau}{=}\frac{Ts}{6}$.

In our scenario, pilots of size $P{=}16$ are transmitted with a comb size $K_c{=}16$. The erroneous AoA from the sensing information are refined using SWOMP with a dictionary matrix considering $d_{\theta}{=}500$. Further, the channel gains $\balphah$ are estimated using SBL considering $d_{\tau}{=}50$. The channel estimation procedure with erroneous sensing information is denoted by $\textbf{SWOMP-SBL\ +\ Sensing\ Info\ Error}$ in the plot. The channel gains $\balphah$ are also estimated using LS assuming perfect sensing information is available at the gNB denoted by $\textbf{Ideal\ Sensing\ Info\ +\ LS}$. 

The performance of our proposed algorithms is evaluated by comparing the normalized mean squared error (NMSE) of the channel using fewer $P$ pilots to the NMSE of the channel obtained by transmitting all the $K=256$ pilots (wideband). The channel estimation using the wideband pilots is performed using the classical LS method and  SWOMP denoted as $\textbf{WB+LS}$ and $\textbf{WB+SWOMP}$ respectively. The dictionary used for both algorithms is of size 500, discretized in the angular domain in $[0^{\circ},180^{\circ}]$.



\begin{figure}[t]
\centerline{\includegraphics[width=3.8in,height=2.4in]{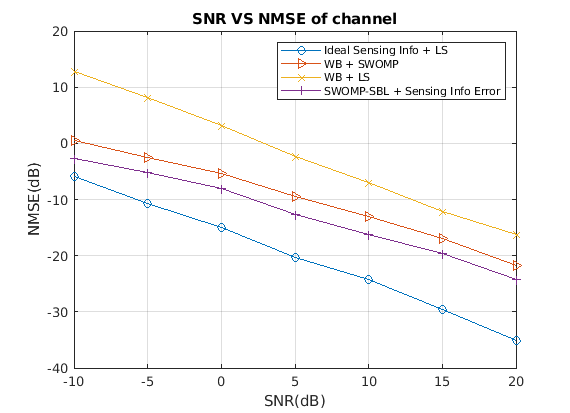}}\vspace{-2mm}
\caption{SNR vs NMSE of the channel estimates in the case of point scatterers.}
\label{fig:case1}
\end{figure}

We consider a scenario where all the scatterers provided by the sensing information in $\mathcal{S}_r$ might not be associated with the communication channel. The parameters of the scatterers chosen for the simulations are $L_r{=}10$ and $L_c{=}6$. From Fig.~\ref{fig:case1}, we can see that with sensing information, SWOMP-SBL based channel estimation algorithm has a significant gain in the NMSE compared to the wideband classical LS and greedy SWOMP algorithm with fewer pilots and robust to the errors in the sensing information. Hence, we reduce the pilot overhead from $100\%$ to $6.25\%$.

\vspace{-2mm}

\section{Conclusion}

In this paper, the uplink channel estimation aided by sensing information for mmWave MIMO systems has been studied. The proposed SWOMP-SBL algorithm, along with the sensing information, uses fewer uplink pilots compared to conventional state-of-the-art systems. The proposed scheme is also robust to erroneous sensing information, including unassociated paths in the sensing information. Simulation results have validated the superior performance using reduced uplink pilots for the proposed SWOMP-SBL scheme compared to conventional state-of-the-art algorithms. Finally, the CRB for the unknown parameters is derived, and local identifiability analysis has been presented.

\bibliographystyle{IEEEtran}\def\baselinestretch{0.9}
\bibliography{refs}
\end{document}